\documentstyle[preprint,aps,prl,overcite]{revtex}
\newcommand{\bea}{\begin{eqnarray}}
\newcommand{\eea}{\end{eqnarray}}
\def\aprle{\buildrel < \over {_{\sim}}}
\def\aprge{\buildrel > \over {_{\sim}}}
\begin{document}
\draft



\title{ 
Relaxation of the cosmological constant at inflation?
}

\author{V.A. Rubakov
}

\address{Institute for Nuclear Research of the Russian Academy of Sciences,\\
         60th October Anniversary prospect, 7a,\\
         Moscow, 117312, Russia;\\
         and\\
         University of Cambridge,\\
         Isaac Newton Institute for Mathematical Sciences,\\
         20 Clarkson Road, Cambridge, CB3 0EH, U.K. 
}

\maketitle
\begin{abstract}
We suggest that the cosmological constant has been relaxed
to its present, very small value during the inflationary
stage of the evolution of the Universe. This requires relatively
low scale, very long duration  
and unconventional source of inflation. We present 
a concrete mechanism of the cosmological constant relaxation at
the inflationary epoch.

\end{abstract}


1. The cosmological constant problem is one of the most challenging
problems in fundamental physics (for a review see, 
e.g., Ref.~\cite{Weinberg}). It would be natural to estimate,
on dimensional grounds, that the vacuum energy density
$\epsilon_{vac}$ is of the order of $M_{Pl}^4$. Supersymmetry
may help to reduce this estimate by many orders of magnitude,
but even the QCD contribution to $\epsilon_{vac}$, which is
of the order of $\Lambda_{QCD}^4 \sim 10^{-3}$ $\mbox{GeV}^4$, is much
greater than the observationally allowed value
$\epsilon_{vac} \aprle 10^{-47}$ $\mbox{GeV}^4$. It is hard to
imagine any symmetry that would ensure (almost) zero present value of
$\epsilon_{vac}$ (what symmetry can possibly take care of the details
of the structure of QCD vacuum?), so it is natural to search for
dynamical explanations of this huge
discrepancy. Among the latter, most appealing would be a
mechanism that would lead to the 
relaxation of the cosmological constant
from fairly arbitrary value towards zero in the course of the
evolution of the Universe.

To the best of author's knowledge, no relaxation mechanism 
close to be successfull has been suggested so far; there even
has been formulated the corresponding 
``no-go theorem''~\cite{Weinberg}. Existing attempts (see 
Refs.~\cite{Dolgov:1997zg,RT} for recent discussion and 
Ref.~\cite{Weinberg} for an account of earlier works) are
grossly inconsistent with Newtonian gravity, as they lead to
exceedingly large values of the effective Planck mass. Besides the
requirement of consistency with Newtonian gravity, there are other
constraints that make the problem difficult. Namely, the
theory of primordial nucleosynthesis requires that much of the
vacuum energy density was already absent at the nucleosynthesis 
epoch, and also that the effective gravitational constant at that
epoch was the same as today to about 10 per cent accuracy.
Thus, the relaxation of the cosmological constant should have
occured, at least partially, at some earlier cosmological
stage. The theory of structure formation in the Universe, that
requires long matter dominated epoch, also points in the same 
direction.

On the other hand, when trying 
to invent a relaxation mechanism operative
at the radiation dominated era, 
one faces the problem of what was special about
{\it vacuum} energy density at that time. At first glance,
the difference between the energy-momentum tensors of vacuum and
radiation is that $T^{(vac)\mu}_{\mu} \neq 0$, so one might
wish to consider the relaxation of $T^{\mu}_{\mu}$. However,
the trace of energy-momentum tensor of relativistic matter
did {\it not} vanish in the early Universe because of interactions
between particles, so the relaxation of  $T^{\mu}_{\mu}$ to zero at
the radiation dominated stage would not mean the relaxation of 
$\epsilon_{vac}$ to an acceptable value.

Although these observations do not necessarily rule out
other options, they suggest that the relaxation of the 
energy density of vacuum of  conventional fields may have 
occured during an inflationary epoch. Such a scenario requires,
of course, some non-standard mechanism of inflation, in which
inflation is driven 
{\it not} by a scalar field, inflaton, 
with conventional properties. 
Once this exotic possibility 
is accepted, the problem to understand what is special about 
vacuum energy density disappears: vacuum is the only component
of conventional matter that does not get inflated away.
Hence, a possible scenario is that during the inflationary stage,
the energy density of the vacuum of conventional fields relaxes
 to (almost) zero, whereas the gravitational ``constant'' (and, maybe,
other coupling ``constants'') settles down to its present value;
these are frozen at later stages, so the
post-inflationary evolution proceeds in the standard way.

This scenario in several respects resembles the
pre-Big-Bang scenario of Ref.~\cite{Veneziano} . Unlike the latter,
however, the relaxation of the cosmological constant needs 
low scale of inflation, for the following reason.
The quantity that one wishes to be realxed to (almost) zero
during the 
inflationary stage is the energy density of the
{\it present day} vacuum. Hence, to a very good accuracy the vacuum 
of conventional matter must be the same at the inflationary
stage as it is today. This requires sufficiently low
Gibbons--Hawking temperature, $T_{GH} \sim H$, where $H$ is
the Hubble parameter at inflation. Almost certainly,
$T_{GH}$ must be much smaller than the QCD scale, and presumably it
must be well below the electron mass. 
Taking, as a crude estimate, $T_{GH} < 10^{-4}$ $\mbox{GeV}$ and writing
$H \sim M_{infl}^2/M_{Pl}$, where $M_{infl}$ is the energy scale
of inflation, one obtains $M_{infl} < 10^{7}$ $\mbox{GeV}$. In fact,
a particular mechanism presented below may require, depending
on parameters, even 
lower scale of inflation.
In this, and a number of other respects, our scenario is similar to
brane-Universe one~\cite{brane-universe} ; in fact, the brane Universe
picture may turn out to be a natural framework beyond our
phenomenological approach.

As far as the relaxation itself is concerned, we propose to make use of
the observation made in the context of
hyperextended inflation   
\cite{Accetta} that singular kinetic terms
of scalar fields
tend to terminate the evolution of 
these fields. This freezing out may occur at  values where
the scalar potential has non-vanishing slope; we will see 
that in a class of models the fields
freeze out in such a way that the value of the scalar potential,
with the
energy density of the vacuum of conventional fields included,
is indeed very small.

Any model with the above properties will be
clearly rather complicated, and will invoke several fields
absent in the Standard Model and many of its extensions.
At the very least, such a model  provides a counter-example
to the no-go theorem of Ref.~\cite{Weinberg} ; optimistically,
it may reflect interesting physics beyond
(almost) zero cosmological constant. It is encouraging that the
energy scales involved are necessarily much smaller than the
Planck scale.

2. As a concrete example, let us consider a model in which the
gravitational interactions of conventional matter fields are of
scalar-tensor type, $\varphi$ being the Brans--Dicke scalar field.
We will need another scalar field $\chi$ with 
the scalar potential $V(\chi)$ and $\varphi$-dependent kinetic term,
and we also include
an inflaton sector. In the Einstein conformal frame, the 
Lagrangian of this model is
\bea
  L = &-&\frac{1}{16\pi G_0} R \sqrt{-g} 
        \nonumber \\
        &+& L_{conv}(\psi;V(\chi);A^2(\varphi)g_{\mu\nu}) 
       \nonumber \\ 
       &+& (L_{k,\varphi} + L_{k,\chi} + L_{infl}) 
\label{lagrangian}
\eea
Here $G_0$ is the present value of the gravitational constant,
$\psi$ stands for all conventional matter fields and
$A^2(\varphi)$ is a conformal factor which is assumed to be
positive at all $\varphi$. The Brans--Dicke 
field  
has canonical kinetic term, 
$L_{k,\varphi} = (1/2)\sqrt{-g}g^{\mu\nu}\partial_{\mu}\varphi
\partial_{\nu}\varphi$, and is defined in such a way that the
Einstein gravity is restored at $\varphi=0$. At this value of
$\varphi$ the conformal
factor $A^2(\varphi)$ is equal to 1, and we assume that in the
vicinity of this point $A^2(\varphi)$ has the form
\begin{equation}
   A^2 = 1 - \frac{1}{2\mu^2}\varphi^2
\label{4n*}
\end{equation}
where $\mu$ is a parameter of dimension of mass. We take
 $\mu \sim M_{Pl}$, as usual 
in scalar-tensor theories,  and will not need to
further fine tune this parameter. These properties of $A^2(\varphi)$, i.e., 
its positivity at all $\varphi$, the absence
of a linear term near $\varphi=0$ and negative $A''(0)$, will
be important in what follows.

The
scalar potential of the field $\chi$ enters $L_{conv}$ in the usual 
way, but the kinetic term
$L_{k,\chi}$ is unconventional,
\[
   L_{k,\chi} = 
   \frac{1}{2} F(\varphi)\partial_{\mu}\chi \partial^{\mu}\chi
\]
The relaxation mechanism is based on the assumption that
$F(\varphi)$ is singular at $\varphi=0$,
\begin{equation}
   F(\varphi) = \frac{\mu^{2p}}{\varphi^{2p}} \,\,\, \mbox{at}\,\,\, 
   \varphi \to 0
\label{5n**}
\end{equation}
with some integer 
exponent $p$ (a numerical coefficient here is a matter
of normalization of $\chi$). In what follows we take
$p \geq 2$. At the moment it is entirely unclear whether fields with
such exotic kinetic terms may have natural particle physics 
interpretation.

Finally, $L_{infl}$ describes an inflaton sector. We require that (i)
the inflaton sector produces inflation with small enough Hubble
parameter $H$, (ii) inflaton energy-momentum tensor (almost) vanishes
today, and (iii) $L_{infl}$ does not contain the fields $\varphi$ 
and $\chi$. 

The properties (ii) and (iii) are rather problematic in
the case of the usual, potential-driven inflation. Indeed, the property 
(ii) implies that the inflaton potential is zero at its minimum;
the mechanism of relaxation to be described below works for
the energy density of the vacuum of
conventional fields only, and does not work for the
inflaton sector. So, one has to assume some symmetry ensuring this
property. The property (iii) means that the inflaton field is indeed
unconventional: the Einstein-frame metrics enters $L_{infl}$ on
its own, and not in the combination $A^2(\varphi)g_{\mu\nu}$. 
We note in this regard that matter which interacts unconventionally
with metrics and Brans--Dicke field has been discussed from a 
different point of view in Ref.~\cite{Gibbons} , and that
such matter (bulk fields) appears in effective four-dimensional 
descriptions of brane world.

The properties (ii) and (iii) seem more natural if inflation is
driven by higher-order terms in the gravitational action
\cite{Starold}. Also, the property (ii) is inherent in
models of $k$-inflation \cite{Mukh}.

3. Let us now consider the behavior of the system at the 
inflationary epoch. As the matter particles have
been inflated away, $L_{conv}$ effectively reduces to
$[-\epsilon_{vac} - V(\chi)]A^4(\varphi)\sqrt{-g}$ 
at this epoch, where $\epsilon_{vac}$
is the energy density of vacuum of conventional fields. 
As discussed above, this vacuum at the inflationary epoch is the
same as today, so our aim is to see whether
$V_{eff}(\chi) = [\epsilon_{vac} +V(\chi)]$ relaxes to a very small
value. We assume that $V_{eff}(\chi)$ takes both positive and negative
values, depending on $\chi$, and consider initial conditions with
$V_{eff}(\chi) > 0$.  

For a very
wide class of initial data, the field $\varphi$ at
the beginning undergoes fast non-linear oscillations, whereas $\chi$
slides along the potential $V_{eff}$. To see this, let us write
the equations for homogeneous scalar fields,
\begin{equation}
  \frac{d}{dt}\left(F(\varphi)\frac{d{\chi}}{dt}\right) 
   + 3HF(\varphi)\frac{d\chi}{dt}
      = - A^4(\varphi)\frac{\partial V_{eff}}{\partial \chi}
\label{B21*}
\end{equation}
\begin{equation}
  \frac{d^2\varphi}{dt^2} + 3H\frac{d\varphi}{dt} =
     -\frac{\partial A^4}{\partial \varphi} V_{eff}
     +\frac{1}{2}\frac{\partial F}{\partial \varphi} 
      \left(\frac{d\chi}{dt}\right)^2
\label{B21**}
\end{equation}
During the initial stage, the Hubble damping is
negligible, and eq.(\ref{B21*}) implies that
$\dot{\chi} \sim f_1(t) F^{-1}(\varphi)$ where $f_1$
is a slowly varying function of time. Equation (\ref{B21**})
is then an equation for a particle with coordinate $\varphi$
in a potential $[f_2(t) A^4(\varphi) + (f_1^2(t)/2) F^{-1}(\varphi)]$
where $f_2$ is another slowly varying function. From eqs.(\ref{4n*})
and (\ref{5n**}) one finds that the latter potential behaves near
$\varphi=0$ as $[-f_2 \mu^{-2} \varphi^2 + (f_1^2/2) \varphi^{2p}
+ \mbox{const}]$. Under mild assumptions about the behavior of
$A^2(\varphi)$ and $F(\varphi)$ at large $\varphi$, this
potential increases towards $|\varphi| \to \infty$, so the
Brans--Dicke field does not run away to infinity.
Depending on parameters and
initial data, $\varphi$ indeed oscillates either about
zero or about a non-vanishing value.

These oscillations are damped because of the expansion of the Universe,
and after several Hubble times the slow roll regime sets in.
The field $\chi$ rolls down the potential $V_{eff}(\chi)$, whereas
$\varphi$ moves towards $\varphi=0$ without oscillations.
If $V_{eff}(\chi)$ is initially large, it dominates at the first 
stage of inflation. Ultimately $V(\chi)$ becomes relatively small, 
inflation becomes driven by $L_{infl}$,  and the Hubble parameter
$H$ becomes approximately constant
and independent of $\varphi$ and $\chi$.
Let us consider explicitly the final stages of
the evolution of $\varphi$ and $\chi$, 
at which $V_{eff}(\chi)$ approaches zero (being
initially positive), and $\varphi$ is close to zero. 
In a neighbourhood of
the point at which $V_{eff}(\chi)=0$, the potential may be approximated 
by  a linear function; by redefining $\chi$ (in a way that depends on
the value of $\epsilon_{vac}$) we set $V_{eff}(\chi)= r\chi$, where
the slope $r = V'$ is positive and has dimension
$(\mbox{mass})^3$. Again, we will not need to fine tune $r$.

In the slow roll approximation, which is very good at the stage we 
discuss, the field
equations at small $\varphi$ and $\chi$ are
\begin{equation}
    3HF\dot{\chi} = -r
\label{32+}
\end{equation}
\begin{equation}
   3H\dot{\varphi} = \frac{2r}{\mu^2}\chi\varphi - 
     \frac{p\mu^{2p}}{\varphi^{2p+1}} \dot{\chi}^2
\label{32*}
\end{equation}
Let us first consider the case $p>2$. We will see that the
relaxation of the vacuum energy density requires fairly small
$H$. Under this assumption,
the fields for long time 
follow the power-law attractor solution,
\begin{equation}
  \chi=\frac{1}{p-1}\left[\frac{p(p-1)}{2}\right]^p
       \frac{r}{(3H)^2}\frac{1}{(3Ht)^{p-1}}
\label{B23*}
\end{equation}
\begin{equation}
  \frac{\varphi^2}{\mu^2}=\frac{p(p-1)}{2} \frac{1}{3Ht}
\label{33*}
\end{equation}
This solution is valid until $\chi$ gets very close to zero.
For this solution, the left hand side of eq.(\ref{32*}) is
negligible, and the two terms on the right hand side cancel each
other. 

The regime (\ref{B23*}), (\ref{33*}) 
terminates when the left hand side of
eq.(\ref{32*}) becomes comparable to $r\chi\varphi/\mu^2$. This occurs at
the time determined by $(3Ht)^{p-2} \sim r^2/[(3H)^4 \mu^2]$.
At this time the effective vacuum energy density $V_{eff}(\chi) = r\chi$
is of order
\begin{equation}
   \epsilon_{*} = \frac{r^2}{(3H)^2} \delta^{p-1}
\label{34*}
\end{equation}
where 
\[
  \delta = \left[\frac{(3H)^4 \mu^2}{r^2}\right]^{\frac{1}{p-2}}
\]
is dimensionless and small at small $H$. The evolution at later times
is more complicated. The field $\chi$ slightly overshoots the point
where $V_{eff}=0$, so that the effective vacuum energy density
becomes {\it negative}. Then the two terms on the right hand side
of eq.(\ref{32*}) work in the same direction and push $\varphi$
towards zero. The dynamics thus freezes out. The final value of
$V_{eff}$ and the relevant time scale become clear after rescaling,
$\chi = r(3H)^{-2} \delta^{p-1}\cdot \tilde{\chi}$, 
$\varphi = \mu\delta^{1/2} \cdot \tilde{\varphi}$,
$t = (3H)^{-1} \delta^{-1}\cdot \tilde{t}$.
Written in terms of variables $\tilde{\chi}$, $\tilde{\varphi}$
and $\tilde{t}$, equations (\ref{32+}) and (\ref{32*}) do not
contain any parameters. Hence, the final value of $\tilde{\chi}$
is of order 1, and the residual vacuum energy density is of
order $\epsilon_{res} \sim -\epsilon_{*}$, where $\epsilon_{*}$
is given by eq.(\ref{34*}). [The property that $\tilde{\chi}$
is finite at $t\to\infty$ can be seen by omitting the second term on
the right hand side of eq.(\ref{32*}), which only diminishes the
final value of $|\tilde{\chi}|$; then eqs.(\ref{32+}) and
(\ref{32*}) are straightforward to solve explicitly.
Needless to say, all above properties are straightforward to check
by numerical calculations.] 

At $p=3$ and $p=4$ 
the residual vacuum energy density $\epsilon_{res}\sim -\epsilon_{*}$
is naturally very small. 
Indeed, at $p=4$ one has $\epsilon_{*} = (\mu^3/r)(3H)^4$
which is of order $H^4$ for $r\sim \mu^3$.
With 
$H\sim M_{infl}^2/M_{Pl}$, this is consistent with the observational 
bound provided that the energy scale of inflation is sufficiently
low, $M_{infl} \aprle$ (a few) TeV. At $p =3$ the residual
vacuum energy density is suppressed even stronger,
 $\epsilon_{*} = (\mu^4/r^2)(3H)^6$, so our relaxation
mechanism is consistent with larger scales of inflation.

It is worth noting that the relaxation of the vacuum energy 
density to its present, very small value occurs only if the
inflationary stage lasts very long. 
The above 
scaling argument implies that the  time scale of the
relaxation is of order
$ H^{-1} \delta^{-1}$, so the duration of inflation should be
large enough, $t_{infl} \aprge  H^{-1}\delta^{-1}$.
On the other hand, the estimate for the residual vacuum
energy, eq.(\ref{34*}), can be written also as
$|\epsilon_{res}| \sim H^2 \mu^2 \delta$. Requiring that 
$|\epsilon_{res}| \aprle \rho_{crit} \sim H_0^2 M_{Pl}^2$
where $H_0$ is the present value of the Hubble parameter,
one finds at $\mu \sim M_{Pl}$ that $\delta \aprle H_0^2/H^2$,
and
\begin{equation}
   t_{infl} \aprge \frac{H}{H_0} t_0
\label{addd*}
\end{equation}
where $t_0 \sim H_0^{-1} \sim 10^{10}$ yrs. Thus, the relaxation
mechanism works only if inflation lasts many orders of
magnitude longer than the entire post-inflationary evolution of 
the Universe. This bizarre requirement is of course a reflection of
the extraordinarily small residual value
of the vacuum energy density.

The very large number of inflationary
e-foldings, $n_e \sim (Ht_{infl}) \sim \delta^{-1}$, 
ensures also  that
$\varphi$ gets very close to zero by the end of inflation,
so that $A^2(\varphi)$ 
does not evolve at later stages (provided that 
$\mu \sim M_{Pl}$), and  interactions 
of the Brans--Dicke field $\varphi$
with
matter are weak enough to satisfy numerous constraints~\cite{Damour}

The case $p=2$ is even simpler to treat. At relatively small $H$
the attractor solution (\ref{B23*}), (\ref{33*}) (with multiplicative
corrections of order $[1 + O(H^4\mu^2r^{-2})]$ ) describes the
evolution of $\varphi$ and $\chi$ all the way to the end
of inflation. The residual
vacuum energy density 
is {\it positive}
in this case, and is determined mostly by the number of
inflationary e-foldings, $\epsilon_{res} \sim (r/3H)^2 n_e^{-1}$.
Hence, at $p=2$ our compensation mechanism is not particularly
sensitive to the scale of inflation, but needs a very large number of
e-foldings. The estimate (\ref{addd*}) holds at $p=2$ as well.

4. The mechanism just described is capable to relax the cosmological 
constant to a very small, but non-zero
value. 
This value may be either negative ($p>2$) or positive ($p=2$).
We note in passing that positive cosmological constant
is obtained also at $p>2$ if inflation terminates when the fields
$\chi$ and $\varphi$ still evolve in the attractor regime
(\ref{B23*}),  (\ref{33*}). It may seem encouraging, 
in view of observational data (see Ref.~\cite{review} 
and references therein), that non-zero and positive cosmological 
constant comes out naturally in our scenario; the expectation then
is that the cosmological constant is time-independent after 
inflation until and long after the present epoch (it is
straightforward to see that, if parameters are not fine tuned,
the evolution of the field $\chi$ is negligible at
post-inflationary stages).
The problem, however,
is that there does not seem to be any chance to address in this 
context
the issue of cosmic coincidence (why $\Omega_{\Lambda}$ is
presently of the order of $\Omega_{Matter}$, and not much
greater or much smaller than $\Omega_{Matter}$ ?). Hence, the
relaxation of the cosmological constant at inflation is a 
less attractive possibility if the cosmological constant is
indeed non-zero today.

The author is indebted to F.~Bezrukov for discussions and
help in numerical 
calculations and to  B.~Bassett, E.~Copeland, A.~Dolgov,
A.~Guth, V.~Lukash, L.~O'Raifeartaigh, M.~Sasaki, M.~Sazhin, 
P.~Steinhardt, P.~Tinyakov and N.~Turok for discussions at various 
stages of this work. This research is supported in part by 
Russian Foundation for Basic Research, grant 990218410.

\end{document}